\documentclass[superscriptaddress,prb,super=false,
showpacs,floatfix,fleqn,longbibliography]{revtex4-1}

\setcitestyle{numbers,square}

\usepackage[dvips]{graphicx}
\usepackage{epsfig} 
\usepackage{subfig}
\usepackage{color}
\usepackage[normalem]{ulem}
\usepackage{amsmath,amsfonts,amssymb,bm}
\UseRawInputEncoding

\begin{document}
\title{How alignment controls heat transport in low dimensional molecular systems  with kinks?}
\author{Igor V.  Parshin}
\author{Igor V.  Rubtsov}
\author{Alexander L.  Burin}
\email[]{aburin@tulane.edu}
\affiliation{Department of Chemistry, Tulane University, New
Orleans, LA 70118, USA}
\date{\today}

\begin{abstract}
Thermal transport in polymer chains is commonly attributed to ballistic propagation of long-wavelength acoustic phonons, which act as Goldstone modes protected by translational symmetry, whereas transport by higher-frequency phonons is suppressed by Anderson localization. Consistent with this picture, highly aligned polymers exhibit exceptionally high thermal conductivity, while poorly aligned polymers are orders of magnitude less conductive and serve as efficient thermal insulators. Here we show that this striking sensitivity to molecular alignment originates from acoustic-phonon scattering by molecular kinks. In the long-wavelength limit, longitudinal acoustic (LA) phonons are completely reflected by a single kink, whereas transverse acoustic (TA) phonons exhibit a universal transmission coefficient of one half. We show that the strong reflection results from the breaking of translational symmetry caused by the change in molecular-axis direction at the kink, while the universal TA transmission originates from virtual scattering through an evanescent transverse Bloch mode. The resulting strong suppression of long-wavelength phonon transport dramatically reduces the thermal conductivity of poorly aligned chains. These findings identify kink engineering as a promising strategy for controlling thermal transport in polymeric materials.
\end{abstract}

\maketitle

\section{Introduction}

Understanding heat transport in disordered molecular chains remains a longstanding challenge.   
The heat transport is governed by two competing mechanisms.   Disorder-induced Anderson localization suppresses vibrational transport in one dimension \cite{Abrahams1979ScThLoc},  whereas ballistic propagation of long-wavelength acoustic phonons acting as Goldstone modes \cite{1962GoldstoneSymmPart}  enhances it \cite{1941PomeranchukDivThCond}.  
The interplay between these effects has been analyzed theoretically in Refs. \cite{Lebowitz67,1998LepriAnomalThCondFPUCl,20011DGlass,Turbulence02,2005a033ntanh} (see also the reviews \cite{LEBOWITZ00,DharReview,2021RevLiviPedagog} and references therein), leading to the prediction of superdiffusive heat transport governed by a narrow band of ballistically propagating low-frequency phonons.   As a result, the thermal conductivity increases with chain length, in contrast to the length-independent conductivity expected in the diffusive regime. This prediction is consistent with numerous experimental observations in molecular systems \cite{MeierMolChains14,2024prlnanopartonpolymsuperdiff}, nanotubes \cite{2008cntLDep,2017cntHighKappa_mmlong,2017cntBencMark},  nanowires \cite{2017BallTr4KSiNanow033,2021SiNanowiresRev} (see also the reviews \cite{2012RevAnCondEurPhZ,2020RevSizeDepTr,2023RevNonFourPhTranspLiviLepri,2024SubSuperDiffTranspThRectif} and references therein),  nanostructures \cite{2000NatQuantThermCond} and even bulk matter \cite{2015SiQuasiBallTransp}.  

In apparent contrast to this picture, molecular dynamics simulations \cite{2019ThermCondKinksXuhui,2020NitzanHeatCondMDRegimes}  reveal a remarkable sensitivity of thermal transport in polymer chains to gauche kinks \cite{1969florystatistical,2003rubinsteinpolymer}, which determine the degree of molecular-chain alignment.   Strongly aligned chains exhibit ballistic transport and thermal conductivity that increases with molecular length, whereas poorly aligned chains display much weaker, nearly diffusive heat transport, resulting in thermal conductivities that are orders of magnitude lower  \cite{1977ChoyReviewPolymerThCond,1988PolymHighThCond,2010PolyethNanoFibHigThCond,2010MDCondSensFluctShort,2012MolChainMDThC,2019PolymerHighThCond,2020MDLangEnergyTrNitzan,2021MDPolyethAnisLiu,2022QuantPhTrDvira,2024YoungModulusAndThermCond}.

To our knowledge, phonon scattering by molecular kinks has not previously been investigated theoretically because the change in molecular-axis direction couples longitudinal and transverse vibrations.   Although the role of transverse vibrations in one-dimensional phonon transport has been demonstrated numerically  \cite{2019ThermCondKinksXuhui,2020NitzanHeatCondMDRegimes,2004TrVibrMD,2005a033ntanh,2008HeatConbdWithTrMod,ab25SuperDiffPh} and in two dimensions (flexural phonons) in Ref.   \cite{2010FlexuralPhononsThCond}, there is no theoretical description of phonon scattering by a molecular kink.  

\begin{figure}%
    \centering
  \includegraphics[width=15cm]{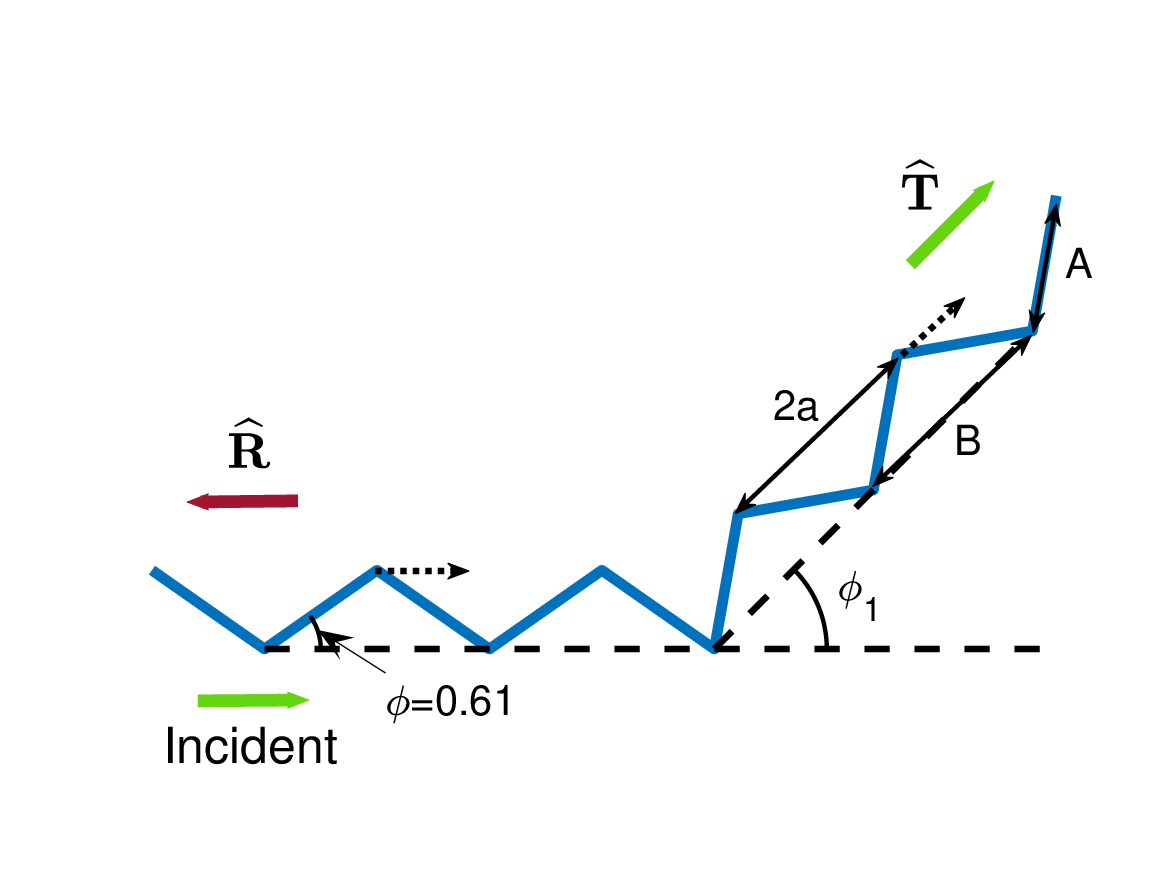} %
    \caption{Models of a kink. Thick solid arrows indicate the directions of the incident, reflected, and transmitted waves, characterized by the reflection and transmission matrices. Dashed lines represent the molecular axes, whereas solid lines denote the interatomic bonds. Dotted arrows illustrate longitudinal displacements associated with the breaking of translational symmetry for longitudinal phonons. The lattice period, force constants, and characteristic angles are indicated as defined in the text.}
    \label{fig:Kink1}%
\end{figure}

Here  we develop a microscopic theory of acoustic-phonon scattering by molecular kinks and investigate its consequences for heat transport.    We show that a change in molecular-axis direction by kinks breaks the translational symmetry that protects Goldstone phonons, creating an anomalous long-wavelength scattering mechanism with profound consequences for thermal transport.  We evaluate the transmission and reflection coefficients of longitudinal acoustic (LA) and transverse acoustic (TA) phonons scattered by a single kink within the fence model of an atomic chain introduced in Sec. \ref{sec:Mod} (see Fig. \ref{fig:Kink1} and Fig.  \ref{fig:Freqvsk}) and reproducing the structure and elastic interactions in  alkane chains  \cite{ab25SuperDiffPh,ab2026KinksAligned}.    Although the calculations are performed within the fence model, the underlying scattering mechanism depends only on the long-wavelength phonon  interactions \cite{landau1986Elasticitytheory} and is, therefore, expected to apply more generally, provided the long-wavelength transverse spectrum retains its quadratic form.

We show that a molecular kink is an anomalously strong scatterer of long-wavelength acoustic phonons (Fig. \ref{fig:KinkTr}).  The change in molecular-axis direction breaks the translational symmetry protecting Goldstone modes, causing the scattering problem to involve an evanescent transverse Bloch mode in addition to the propagating transverse acoustic mode.   As a consequence, longitudinal phonons are completely reflected, whereas transverse phonons exhibit a universal transmission coefficient of one half.  The anomalous scattering originates from the one-dimensional nature of the problem: a kink changes the direction of the molecular axis and thereby breaks the translational symmetry protecting Goldstone phonons.  

Our central finding is that molecular kinks act as anomalously strong scatterers of long-wavelength acoustic phonons.  
Such scattering strongly enhances Anderson localization of low-frequency phonons,  since  in one dimension the localization length is controlled by the elastic mean free path  \cite{Abrahams1979ScThLoc} and therefore by the average spacing between kinks.  Indeed, we observe substantial localization of normal modes in randomly generated chains containing kinks (see Sec. \ref{sec:Res} and Fig. \ref{fig:EigMod}). We further demonstrate that phonon scattering by multiple randomly oriented kinks leads to a rapid decrease in thermal conductivity with increasing chain length, in striking contrast to aligned molecular chains \cite{ab2026KinksAligned}. For long chains, we find that the thermal conductivity of strongly aligned chains exceeds that of poorly aligned chains by almost three orders of magnitude (see Fig. \ref{fig:ThCond}).

Our analysis is restricted to harmonic interactions, an approximation that is well justified for aligned chains at and below room temperature \cite{ab25SuperDiffPh}. While anharmonicity may lead to slow diffusive transport in poorly aligned molecules through phonon hopping between localized states, it does not alter the central conclusion that kink-induced phonon scattering strongly suppresses thermal transport.  Such hopping transport may account for the thermal conductivity observed in poorly aligned polymers commonly used as thermal insulators.

We also restrict our analysis to planar geometry with atomic displacements confined to the molecular plane,   since this geometry is simpler, while it captures the coupling between longitudinal and transverse vibrations.   We expect the principal conclusions to remain valid for three dimensional displacements, where additional torsional modes introduce extra transport channels without altering the underlying symmetry-breaking mechanism.

\section{Model}
\label{sec:Mod}

The fence model consists of a planar harmonic chain whose equilibrium geometry follows the positions of carbon atoms in an alkyl backbone, as illustrated in Fig.~\ref{fig:Kink1} (see Refs.~\cite{ab25SuperDiffPh,ab2026KinksAligned}). The model includes both nearest- and next-nearest-neighbor interactions with force constants $A$ and $B$, respectively. On the one hand, this geometry reproduces the arrangement of carbon atoms in polyethylene reasonably well, making the model physically realistic. On the other hand, it represents the simplest model that simultaneously supports longitudinal and transverse vibrations. Indeed, at least nearest- and next-nearest-neighbor interactions are required to stabilize a planar chain under the assumption that each atom occupies its equilibrium position. In contrast, a purely linear chain requires additional interactions to define bending rigidity, while angular-dependent potentials \cite{2004TrVibrMD} lead to a considerably more complicated Hessian matrix.

The elastic energy is written in a translationally and rotationally invariant form,
\begin{eqnarray}
\mathbf{H}=\sum_{n=-\infty}^{\infty}\left(\frac{\mathbf{p}_n^2}{2M}+\frac{A}{2}((\mathbf{u}_{n+1}-\mathbf{u}_n)\mathbf{e}_{n,n+1})^2+\frac{B}{2}((\mathbf{u}_{n+2}-\mathbf{u}_n)\mathbf{e}_{n,n+2})^2\right),
\label{eq:Ham}
\end{eqnarray}
where $M$ is the site mass, $\mathbf{u}_n$ and $\mathbf{p}_n$ denote the displacement and momentum of site $n$ confined to the chain plane, and $\mathbf{e}_{n,k}$ is the unit vector connecting sites $n$ and $k$. Unless stated otherwise, we use $A\approx 4.8B$ and $B\approx 0.0244$ a.u. ($E_h/a_0^2$), obtained from density functional theory (DFT) calculations of the Hessian matrix for alkyl chains \cite{ab2023Cherenkov}, together with the atomic mass $M\approx 14$ a.m.u. The equilibrium geometry is specified by the bond angle $\phi=0.61$ rad ($35^{\circ}$, see Fig.~\ref{fig:Kink1}), corresponding to an inter-bond angle of $110^{\circ}$, close to that in alkane chains. The chain period is $2a=2.45$~\AA.

We denote $a$ as half of the chain period because the transformation $u_{ny}\rightarrow (-1)^n u_{ny}$ reduces the translational period by a factor of two. For an infinite kink-free chain, propagating acoustic phonons are characterized by wave vector $k$ and frequency $\omega(k)$. The displacement field has the Bloch form $\mathbf{u}(n,t)=\mathbf{u}(k,n)e^{-i\omega(k)t}$, where $u_x(n,k)=e_x(k)e^{ikan}$ and $u_y(n,k)=e_y(k)(-1)^ne^{ikan}$, following Bloch's theorem \cite{Kittel2004} and Ref.~\cite{ab25SuperDiffPh}. The phonon dispersion $\omega(k)$ and polarization vector $\mathbf{e}(k)$ are given given by 
\begin{eqnarray}
\omega(k)^2=\frac{4ABM^{-1}\sin(ka)^2(1+\cos(ka))(1-\cos(2\phi))}{A(1-\cos(2\phi)\cos(ka))+2B\sin(ka)^2)+\sqrt{(2B(\sin(ka/2)^2)+A(\cos(2\phi)-\cos(ka)))^2+A^2\sin(2\phi)^2\sin(ka)^2}},  
\nonumber\\
\begin{bmatrix}
e_{x}(k) \\
e_{y}(k) 
\end{bmatrix} = \frac{1}{\sqrt{(\omega(k)^2-2A\sin(\phi)^2(1+\cos(ka)))^2+A^2\sin(2\phi)^2}} \begin{bmatrix}
\omega(k)^2-2A\sin(\phi)^2(1+\cos(ka/2)) \\
iA\sin(2\phi)
\end{bmatrix}.  
\label{eq:EigMod}
\end{eqnarray}

\begin{figure}%
    \centering
  \includegraphics[width=15cm]{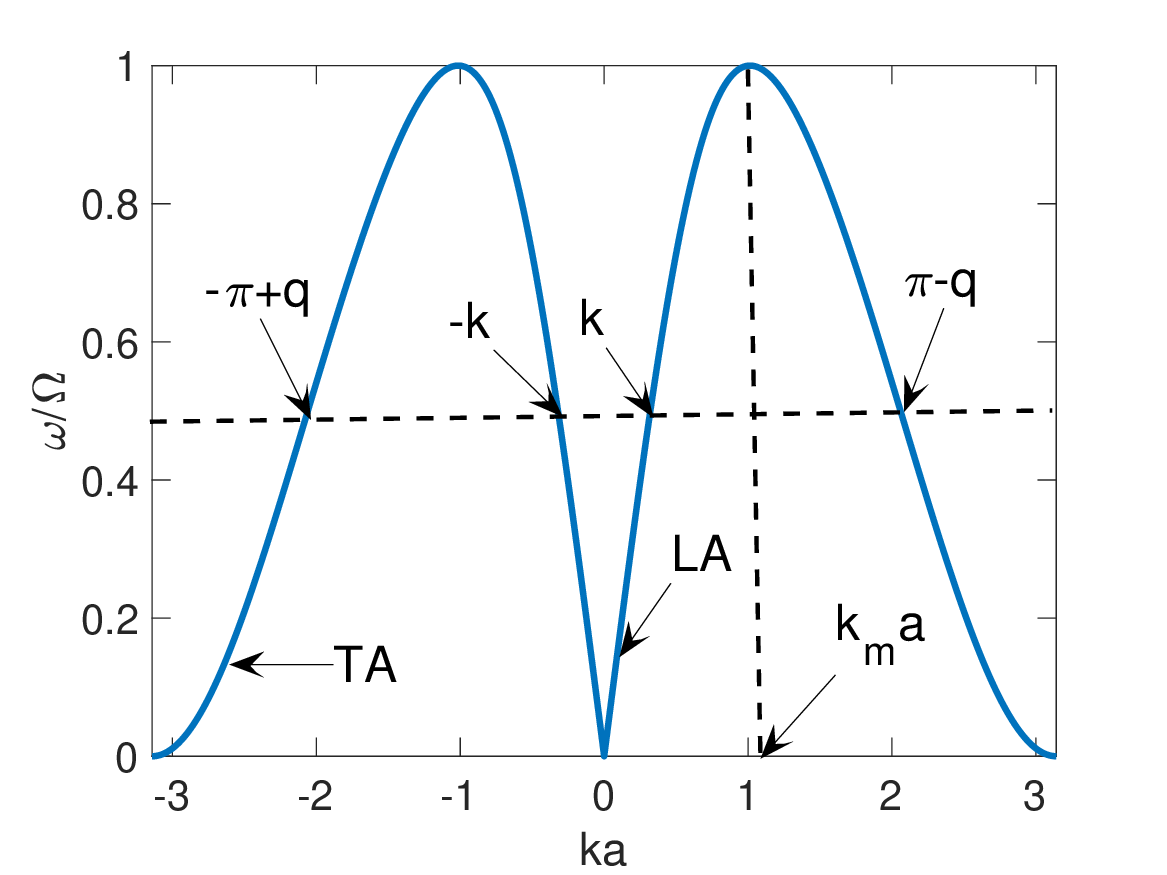} %
    \caption{Spectrum of acoustic waves. The maximum frequency, $\Omega\approx 1.23\sqrt{B/M}$, defines the width of the acoustic band. The wavevector corresponding to the maximum frequency is $k_{\rm m}=1.014/a$. The four solutions of the dispersion equation for a given frequency are indicated by the arrow labels.}%
    \label{fig:Freqvsk}%
\end{figure}

The dispersion is shown in Fig.~\ref{fig:Freqvsk}. The maximum frequency $\Omega\approx 1.23\sqrt{B/M}$ occurs at $k_{\rm m}\approx 1.014/a$ and defines the acoustic bandwidth. For each frequency $0<\omega<\Omega$, there are four real wave vectors $(-\pi+q,-k,k,\pi-q)$ ($0<k<k_{\rm m}$). The central pair $\pm k$ corresponds predominantly to longitudinal modes, while $\pm(\pi-q)$ corresponds predominantly to transverse modes. This distinction becomes exact in the long-wavelength limit, where polarization becomes aligned with either along the molecular chain or perpendicular to it.

In the long-wavelength limit, the longitudinal and transverse dispersions reduce to
$\omega_{\rm l}(k)\approx ck$ and
$\omega_{\rm t}(\pm \pi \mp q)\approx \eta c a q^2$, where $c=2a\sqrt{B/M}$ and $\eta=\tan(\phi)/4$ \cite{ab25SuperDiffPh}. The quadratic transverse dispersion follows from rotational invariance \cite{landau1986Elasticitytheory,1993cntSpectrum}.

One additional remark is in order. For the present fence model, and more generally for one-dimensional systems with rotational invariance, the long-wavelength transverse acoustic branch obeys a quadratic dispersion of the form $\omega^2 = A q^4 + \mathcal{O}(q^6)$. Consequently, the leading-order dispersion relation admits four solutions for the wavevector, obtained from the analytic continuation of the asymptotic form. Two of these correspond to propagating transverse modes $q = \pm \sqrt{\omega/(\eta a c)}$, while the remaining two are purely imaginary, $q = \pm i \sqrt{\omega/(\eta a c)}$, and describe evanescent transverse Bloch waves.

Although these evanescent solutions do not carry energy flux, they arise naturally as part of the complete solution space of the lattice dynamical equations in the long-wavelength regime. In particular, they represent the analytic continuation of the same transverse acoustic branch and are required to construct a complete scattering basis satisfying boundary matching conditions at a localized defect (see Sec.~\ref{sec:Methods}).

In this limit, the imaginary wavevector scale satisfies $Q \sim q$, reflecting the fact that propagating and evanescent solutions originate from the same asymptotic dispersion relation.

A kink is introduced as a local rotation of the molecular axis by angles $\pm\phi_1$. For gauche defects in polyethylene, $\phi_1\approx 45^{\circ}$ \cite{1969florystatistical,2003rubinsteinpolymer}, as illustrated in Fig.~\ref{fig:Kink1}. This rotation modifies the equilibrium positions of atoms and therefore the elastic energy in Eq.~(\ref{eq:Ham}). Unless stated otherwise, force constants near the kink are assumed unchanged, since DFT calculations indicate that their variation is below $20\%$ and has only a minor effect on phonon scattering.  To evaluate reflection and transmission coefficients,  we consider an infinite chain containing a single kink (Fig.~\ref{fig:Kink1}).

Thermal conductivity is studied using long chains containing many kinks. We consider poorly aligned (random-walk) and strongly aligned configurations (Sec.  \ref{sub:ThCond}). In the random-walk case, kinks are placed randomly with an average density of one kink per a chain period and random orientation. Chain intersections are discarded since preliminary estimates indicate their negligible effect on thermal conductivity.  Aligned chains are generated with the same kink density but with transverse displacements restricted to two chain periods, as described in Ref.~\cite{ab2026KinksAligned}. In both cases, initial and final molecular-axis directions are parallel and lie at the same vertical position.

\section{Methods}
\label{sec:Methods}

\subsection{Evaluation of transmissions and reflections coefficients}
\label{sub:TransRefl}

Reflection and transmission matrices were evaluated for an otherwise periodic infinite chain containing one or more kinks located between sites $0$ and $N$ ($N>0$) using the transfer-matrix method developed in Refs.~\cite{ab25SuperDiffPh,ab2026KinksAligned}. We first consider scattering of an incident longitudinal acoustic (LA) phonon with wavevector $0<k<k_{\rm m}$; the case of TA phonons is analogous.

For the left lead ($n<0$), the scattering state consists of the incident wave, two reflected propagating waves, and one evanescent mode. The amplitudes $r_{ll}$ and $r_{tl}$ define the reflection coefficients $R_{\rm ll}=|r_{\rm ll}|^2$ and $R_{\rm tl}=\eta_g|r_{\rm tl}|^2$, where $\eta_g$ is the ratio of group velocities.

For the right lead ($n>N$), the solution consists of two transmitted propagating waves and one evanescent contribution, where polarization vectors are expressed in a basis rotated to align with the right lead with similar definition of transmission  coefficients $T_{\rm ll}$ and $T_{\rm tl}$. 

The transmission and reflection matrices $T_{\mu\nu}$ and $R_{\mu\nu}$ ($\mu,\nu=l,t$) satisfy flux conservation, $\sum_{\nu}(T_{\mu\nu}+R_{\mu\nu})=1$. 

To compute the scattering amplitudes, we introduce a partially rotated Hessian matrix $\widehat{H}_{\rm r}$ obtained by rotating the right semi-infinite lead so that its molecular axis aligns with that of the left lead. In this representation, the system reduces to the perfect-chain Hessian $\widehat{H}_0$ outside the kink region, and the scattering problem is described by the localized perturbation
\begin{eqnarray}
\widehat{V}=\widehat{H}_{\rm r}-\widehat{H}_0.
\label{eq:PertHess}
\end{eqnarray}

Transmission and reflection amplitudes are then computed using the Green-function formalism \cite{1971TrThrGFCaroli,1994MWCoductMR,2004TransmGFRatner,2008QuantThermTranspRev,XingThTrReview18}. In practice, the standard implementation becomes numerically unstable for $\omega\lesssim10^{-3}\Omega$, where flux conservation is violated. Since we are interested in the long-wavelength limit, we developed a modified Green-function approach stable down to $\omega \sim 10^{-6}\Omega$. 

We exploit a spatial partitioning of the degrees of freedom into internal kink coordinates and lead boundary sites. The internal subspace is described by the finite matrix $\widehat{H}_{\mathrm{\rm in}}$, while the coupling to the leads is encoded in matrices $\widehat{V}_\mathrm{l}$ and $\widehat{V}_\mathrm{r}$, which connect the internal domain to the left and right boundary layers, respectively.

The internal Green function is defined as $\widehat{G}_{\mathrm{in}}(\omega)
=\left[\omega^2 \widehat{I} - \widehat{H}_{\mathrm{in}} + i\epsilon \right]^{-1}$.  Eliminating the internal degrees of freedom yields an effective relation between boundary displacements, 
$\mathbf{U}_{\mathrm{in}}=\widehat{G}_{\mathrm{in}} \left(\widehat{V}_\mathrm{l}^\dagger \mathbf{U}_\mathrm{l} +
\widehat{V}_\mathrm{r}^\dagger \mathbf{U}_\mathrm{r}
\right)$,
where $\mathbf{U}_\mathrm{l}$ and $\mathbf{U}_\mathrm{r}$ are vectors collecting the displacement components of the boundary sites of the left and right leads that are directly coupled to the kink region.

Substituting this expression into the equations of motion for the lead boundary sites reduces the scattering problem to a closed system of linear equations for the reflection and transmission amplitudes.  The resulting linear system is solved directly for the scattering amplitudes, from which the transmission and reflection matrices are constructed as described above.

\subsection{Evaluation of thermal conductivity}
\label{sub:methkap}

Thermal conductivity was calculated using the Landauer formula \cite{Landauer57Classic,1988EquatForThermCond,1998QuantThermCond,1998HeatTranspRev,SegalNitzan03,2008TransmDefcnt}, generalized to multichannel transport \cite{1986ImryLandauerMult}. In the high-temperature limit ($k_B T\gg \hbar c/a$) and for small temperature bias, the conductivity of a chain of length $L$ is
\begin{eqnarray}
\kappa = \kappa_{a} \frac{L}{a\omega_{\rm max}}\int_0^{\omega_{\rm max}}d\omega \mathcal{T}(\omega),
\quad
\kappa_{a}=\frac{k_B a\omega_{\rm max}}{2\pi},
\nonumber\\
\mathcal{T}=\mathcal{T}_l+\mathcal{T}_t,
\quad
\mathcal{T}_l=T_{ll}+T_{lt},
\quad
\mathcal{T}_t=T_{tl}+T_{tt}.
\label{eq:kappaDvira}
\end{eqnarray}

For sufficiently long chains ($L>100a$), the high-temperature assumption is not required since transport is dominated by low-frequency phonons \cite{20011DGlass}.

Transmission matrices are computed as described above, and the Landauer integral is evaluated numerically using MATLAB \cite{MATLAB:2019}. Results are averaged over approximately $200$ random realizations ($100$ for the longest chains), yielding statistical uncertainties of a few percent. The code is available in a GitHub repository \cite{abGITHUB}.

\begin{figure}%
    \centering
   \includegraphics[width=15cm]{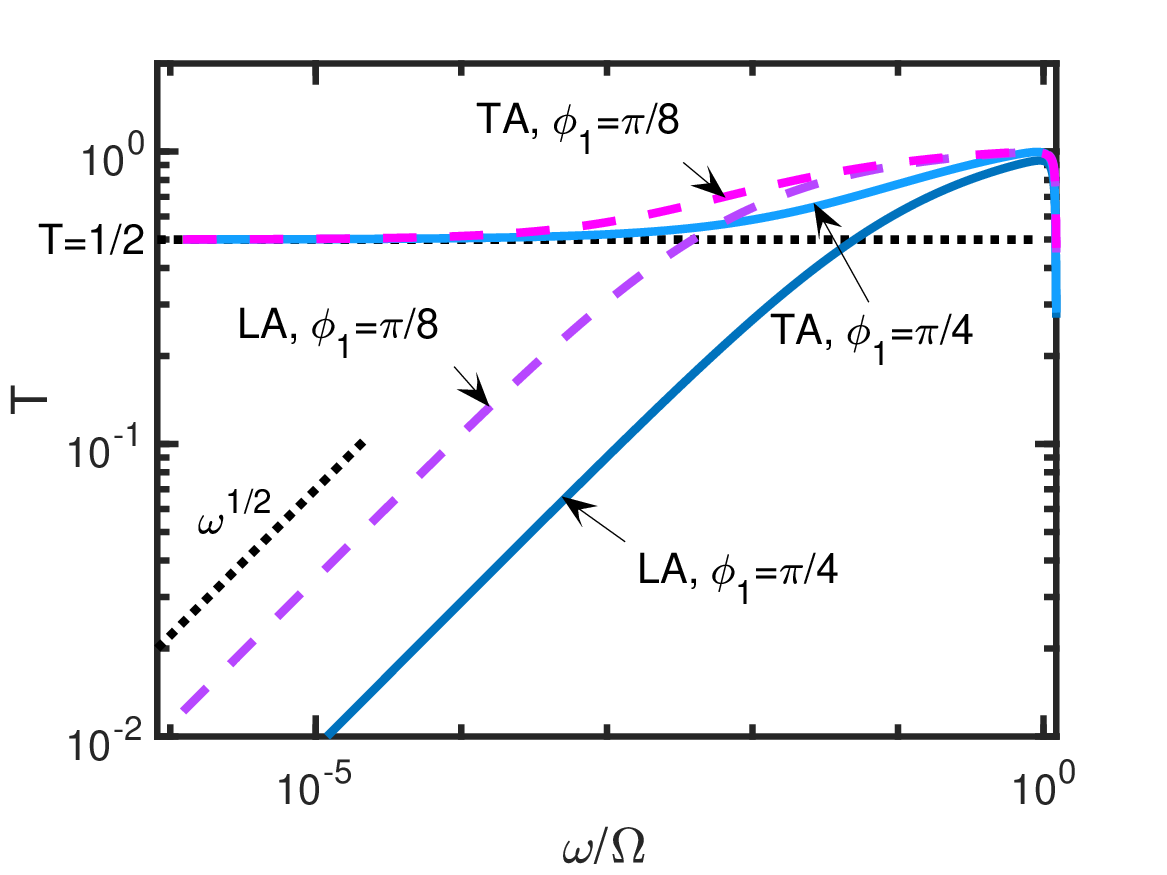} %
    \caption{Transmissions of phonons scattered by a single kink within the fence model.    For the fence model we show the total transmissions  $T_{\mu}=\sum_{\nu}T_{\mu\nu}$ for longitudinal (LA) and transverse (TA) acoustic phonons.    Asymptotic frequency dependencies of transmissions in the long-wavelength limit are illustrated by dotted lines.  }%
    \label{fig:KinkTr}%
\end{figure}


\section{Results}
\label{sec:Res}

\subsection{Phonon scattering by kinks}
\label{sub:KinkSc}

The frequency-dependent transmission coefficients for a single kink are shown in Fig.~\ref{fig:KinkTr}.   Solid lines correspond to the physically relevant kink angle $\phi_1=\pi/4$ (gauche-like geometry), while dashed lines show results for a smaller angle $\phi_1=\pi/8$.

In both cases,  transmissions of LA phonons vanish  in the long-wavelength limit, with transmission scaling approximately as $\omega^{1/2}$. In contrast,  transmissions of TA phonons approach a finite asymptotic value $T_{\rm t}\to 1/2$. This asymptotic behavior is robust across all investigated kink angles $0<\phi_1<\pi/2$.

For small kink angles $\phi_1\ll 1$, the prefactor of the low-frequency LA transmission increases strongly, scaling approximately as $\phi_1^{-2}$, while the $\omega^{1/2}$ dependence is preserved.

The limiting value $T_{\rm t}\to 1/2$ indicates a nontrivial partitioning of transverse energy at low frequencies. Importantly, this result is insensitive, within numerical accuracy, to local modifications of force constants near the kink, including symmetric and asymmetric perturbations of nearest- and next-nearest-neighbor interactions, as well as moderate variations in equilibrium geometry. This suggests that the effect is governed primarily by symmetry constraints and long-wavelength matching conditions rather than microscopic details.

\begin{figure}
\centering
\includegraphics[width=.9\linewidth]{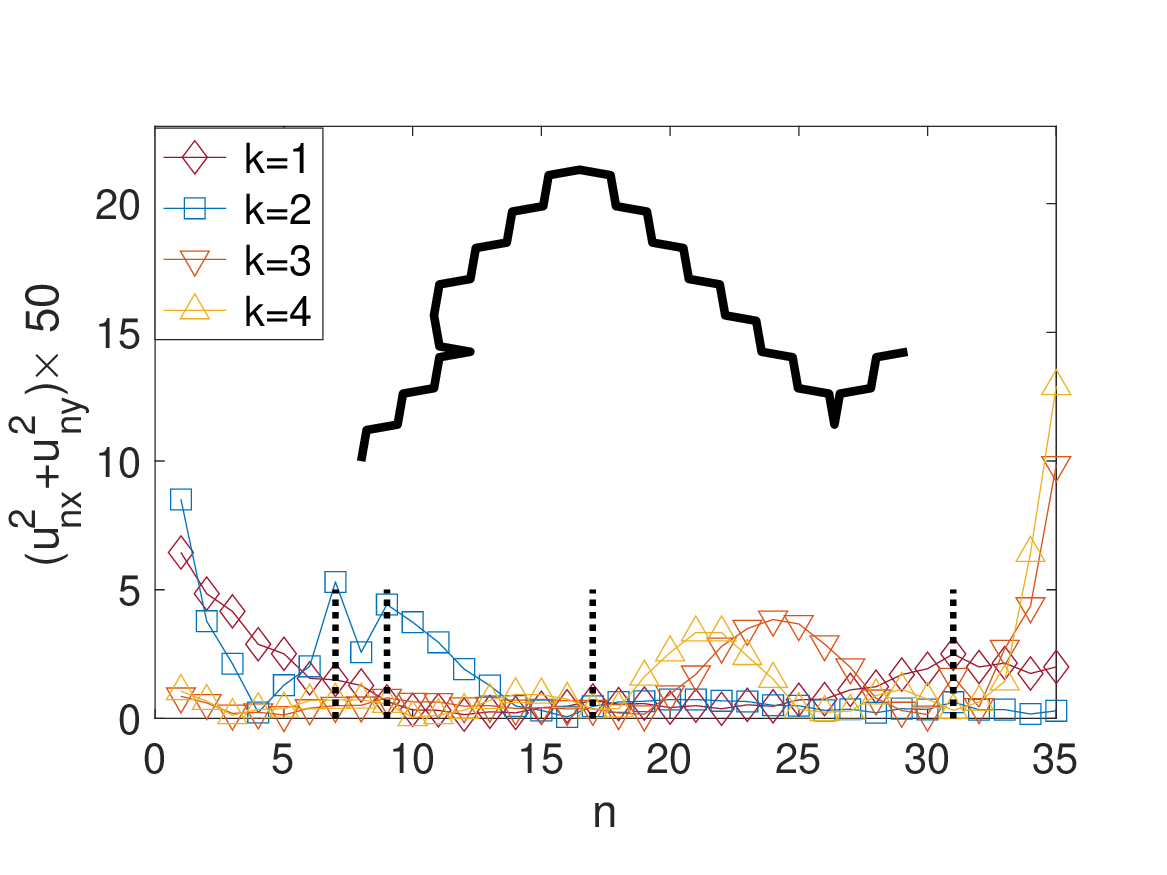}
\caption{Squared vibrational site amplitudes for four lowest-frequency normal modes for the chain shown in the top. Vertical black dotted lines indicate kink positions.}
\label{fig:EigMod}
\end{figure}

To illustrate the consequences for vibrational localization, we diagonalized the Hessian matrix for a finite chain containing four kinks (Fig.~\ref{fig:EigMod}). After excluding three trivial zero-frequency modes (two translations and one rigid rotation), the lowest-frequency vibrational eigenmodes are substantially  localized between neighboring kinks. This is consistent with earlier  molecular-dynamics studies of kinked polymer chains \cite{2019ThermCondKinksXuhui}.

\subsection{Analytical investigation of transmission and reflection within a toy model}
\label{subrestoy}

The following analysis does not aim to provide a complete microscopic description of the kink, but rather to identify a minimal long-wavelength structure that reproduces the numerically observed asymptotic behavior.

We interpret the strong scattering as a consequence of local violation of translational invariance induced by the kink. This is reflected in the structure of the effective Hessian perturbation matrix $\widehat{V}$, Eq.~(\ref{eq:PertHess}), which does not satisfy the translational invariance condition $\sum_{n}V_{mn}=0$.

In the long-wavelength limit, displacement fields vary slowly across the defect, allowing the approximation $u_{n,\mu}\approx u_{0\mu}$. The kink contribution to the energy can then be written in the local form
\begin{eqnarray}
\widehat{v}\approx \frac{k_{\rm l}u_{\rm 0,l}^2}{2}+\frac{k_{\rm t}u_{\rm 0t}^2}{2}.
\label{eq:KinkPinning}
\end{eqnarray}

There is no cross term proportional to $u_{\rm 0,l}u_{\rm 0,t}$ due to the reflection symmetry of the kink. The effective parameters $k_{\rm l}$ and $k_{\rm t}$ depend on $A$, $B$, and the geometry angles $\phi$ and $\phi_1$ (Fig.~\ref{fig:Kink1}). We do not specify them explicitly since they are not required for the qualitative conclusions.

The perturbation in Eq.~(\ref{eq:KinkPinning}) breaks translational invariance and acts as an effective pinning center. In one-dimensional systems, such perturbations generically suppress long-wavelength transport channels \cite{2010LebowitzPinningPot}.

Within the Green function formalism \cite{1971TrThrGFCaroli,1994MWCoductMR,2004TransmGFRatner,2008QuantThermTranspRev,XingThTrReview18}, the dressed propagator in the presence of the kink takes the form
\begin{eqnarray}
G_{\mu}(n,p,\omega)=g_{\mu}(n,p,\omega)+
\frac{k_{\mu}g_{\mu}(n,0,\omega)g_{\mu}(0,p,\omega)}
{1+k_{\mu}g_{\mu}(0,0,\omega)},
\nonumber\\
g_{\rm l}(n,p,\omega)\approx-\frac{i e^{ik|n-p|}}{2\omega v_{\rm l}},\quad
g_{\rm t}(n,p,\omega)\approx-\frac{i e^{iq|n-p|}}{2\omega v_{\rm t}(\omega)}
+\frac{e^{-q|n-p|}}{2\omega v_{\rm t}(\omega)},
\label{eq:GF}
\end{eqnarray}
where $p$ stands for the position of source,   $n$ for the observation point,  $v_{\rm l}\approx c$ and $v_{\rm t}(\omega)\approx \omega/(2q)$ are phonon group velocities.

The transmission amplitude follows from the asymptotic relation
$G_{\mu}(n,p,\omega)= t_{\mu\mu}g_{\mu}(n,p,\omega)$ for $n\to\infty$, $p\to-\infty$. In this limit, the evanescent contribution is negligible in the numerator but remains essential in the denominator. This yields 
\begin{eqnarray}
t_{\rm ll} = 1-\frac{1}{1+\frac{2\omega v_{\rm l}}{ik_{\rm ll}}},  ~~t_{\rm ll} = 1-\frac{1}{1-i+\frac{2\omega v_{\rm t}}{ik_{\rm ll}}}.  
\label{eq:TrEst}
\end{eqnarray}

In the long-wavelength limit $\omega\to 0$, this gives $T_{\rm l}\to 0$ and $T_{\rm t}\to 1/2$, consistent with the numerical results (Fig.~\ref{fig:KinkTr}). The difference between longitudinal and transverse behavior originates from the presence of the evanescent transverse channel, which modifies the analytic structure of the Green function but does not contribute to energy transport.

\subsection{Thermal conductivity at different lengths and alignments}
\label{sub:ThCond}

\begin{figure}
\centering
\includegraphics[width=.75\linewidth]{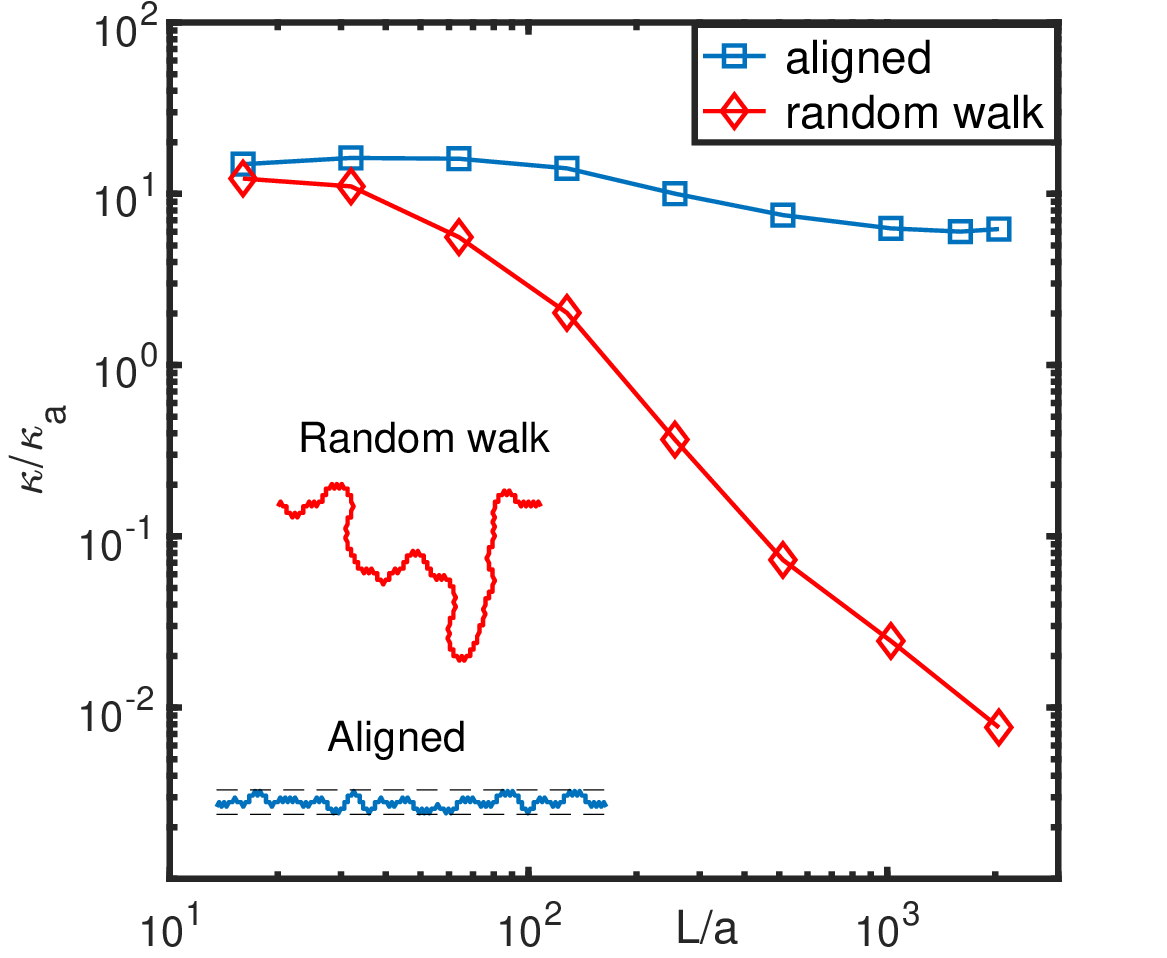}
\caption{Thermal conductivity for random and aligned chain geometries.}
\label{fig:ThCond}
\end{figure}

To quantify the impact of kink scattering on transport, we consider chains containing multiple kinks with fixed turn angles, $\phi_1=\pm\pi/4$, representative of gauche defects in alkane-like systems. Two geometries are studied: (i) random-walk configurations, where kink positions and orientations are uncorrelated, and (ii) aligned configurations, where transverse fluctuations are constrained within one chain period ($2a$) in either direction following Ref.~\cite{ab2026KinksAligned}.
In both cases, kink separations follow a Poisson distribution with mean spacing $l=2a$ (Fig.~\ref{fig:ThCond}).  

The length dependence of thermal conductivity computed using Eq.~(\ref{eq:kappaDvira}) is shown in Fig.~\ref{fig:ThCond}.  The difference of thermal conductivities between the two geometries grows rapidly with the chain length, reaching almost three orders of magnitude for the longest chains considered. This dramatic suppression of thermal conductivity in random-walk chains is consistent with the expectations of kink-induced localization of low-frequency phonons. 

This behavior is consistent with strong multiple scattering of low-frequency phonons by kinks and a progressive suppression of ballistic transport. However, the observed decay is weaker than a purely exponential dependence over the accessible range, suggesting that transport is not fully localized within the simulated system sizes. This may reflect a crossover regime in which very low-frequency modes retain relatively long mean free paths and dominate the residual heat transport.

\section{Discussion}
\label{sec:Disc}

\subsection{Origin of strong reflection of long-wavelength phonons by kinks}
\label{sub:DiscNonvanish}

We find that long-wavelength LA and TA phonons are strongly reflected by molecular kinks, in contrast to conventional structural defects that preserve the molecular axis, where reflection typically vanishes in the long-wavelength limit due to translational invariance \cite{20011DGlass,ab25SuperDiffPh,1962GoldstoneSymmPart}.

A kink modifies this picture by breaking translational invariance separately for longitudinal and transverse displacements. As illustrated in Fig.~\ref{fig:Kink1}, a uniform displacement across the kink changes interatomic distances and therefore produces a finite elastic response even in the long-wavelength limit. This leads to finite reflection for both LA and TA modes, with vanishing LA transmission as $\omega\to 0$ and a finite reduction of TA transmission.

The analytical toy model captures these features by reducing the defect to an effective local perturbation that breaks translational symmetry. Within this description, the difference between LA and TA scattering originates from the presence of an evanescent transverse channel, which contributes to the Green function but does not transport energy. We interpret this as a virtual scattering pathway absent in the purely longitudinal sector.

The robustness of the result with respect to local modifications of the kink region suggests that it is controlled primarily by symmetry and long-wavelength boundary matching rather than microscopic details.

\subsection{Transport in chains with multiple kinks}
\label{sub:DiscManyKinks}

For chains containing multiple randomly oriented kinks, the results in Fig.~\ref{fig:ThCond} show a strong suppression of thermal conductivity with increasing length compared to aligned geometries.

In random-walk configurations, successive kinks act as anisotropic scatterers that disrupt both longitudinal and transverse propagation channels, leading to strong attenuation of coherent phonon transport.

However, the length dependence is weaker than a simple exponential decay over the accessible range, indicating that transport is not fully localized within the simulated system sizes. One possible interpretation is a crossover regime in which the effective mean free path of low-frequency modes increases rapidly as frequency decreases, allowing these modes to dominate residual transport. This is consistent with the presence of three translational zero modes and suggests that the lowest non-zero frequency modes retain partial delocalization.

In aligned chains, partial correlation between successive kinks restores an effective large-scale symmetry, significantly reducing backscattering and enhancing thermal transport.

More broadly, these results highlight that geometry-induced symmetry breaking can play a role comparable to disorder in controlling phonon localization in one-dimensional systems. A complete characterization of the asymptotic transport regime remains an open problem.

\section{Conclusion}
\label{sec:Conclusion}


We investigated phonon transport in molecular chains with kinks within a two-dimensional fence model representing alkane-like backbones. We find that, in contrast to conventional structural defects, phonon reflection by kinks does not vanish in the long-wavelength limit. This behavior is interpreted as a consequence of the violation of translational invariance for longitudinal (LA) and transverse (TA) acoustic phonons induced by changes in the molecular axis.

This mechanism has important consequences for thermal transport in disordered chains. We observe numerically that thermal conductivity is strongly suppressed in poorly aligned chains compared to strongly aligned ones, with differences approaching three orders of magnitude for the longest systems studied. This suppression originates from strong phonon scattering by kinks, while in aligned configurations it is significantly reduced due to partial cancellation of successive scattering events. At the same time, the thermal conductivity does not exhibit a clear exponential decay with length, suggesting that transport is not governed by a single length-independent localization scale over the accessible system sizes. Instead, heat transport in the disordered regime is likely dominated by a small fraction of weakly scattered low-frequency modes.

For a single kink, we find that the transmission of LA phonons vanishes in the long-wavelength limit, while the TA transmission approaches the universal value $T_{\rm t} \to 1/2$. We attribute this asymmetry to the coupling of transverse phonons to evanescent transverse Bloch waves that coexist with propagating modes in the long-wavelength regime due to the structure of the transverse spectrum. This mechanism appears robust with respect to variations of local force constants and geometric details of the kink, suggesting that it may be relevant for real molecular systems. As a result, kinks may act as effective filters that strongly suppress longitudinal transport while allowing partial transmission of transverse modes.  Generally they can serve as significant elements in phononic nanodevices like a thermal diode,  whose performance is otherwise restricted by  the  ballistic long-wavelength phonon transport \cite{2013ThermalDiode}.

Finally, the existence of evanescent transverse Bloch solutions at low frequencies suggests the possibility of localized vibrational states near edges or defect configurations coexisting with extended states at the same frequency. Such coexistence is reminiscent of bound states in the continuum in wave systems. Whether this mechanism can be realized in more realistic molecular settings, and whether it can be exploited for controlled confinement or manipulation of vibrational energy, remains an interesting direction for future work.



\noindent
{\bf Acknowledgement}

This work is supported by U.  S.  National Science Foundation
under Grant No. CHE-2201027 and in part by  U.S. National Science Foundation under Grant Number OIA-2437963 and the Louisiana Board of Regents.

\noindent
{\bf Availability of Data}

\noindent
The data that support the findings of this study are openly available in GitHub at \url{https:
//github.com/aburin25/Thermal-Conductivity-with-Kinks/blob/main/readme.txt}.





\bibliography{pnas-sample}

\noindent
{\bf Author contributions}

\noindent
IP calculated scattering by kink,  IR set up the problem and suggest its applications, AB set up the problem, developed codes for scattering by kinks and thermal conductivity and analytical model, and wrote the text.

\noindent
{\bf Competing interests}

\noindent
The authors declare no competing interests.

\end{document}